\documentstyle[12pt,aaspp4]{article}

\def\ni{\noindent}
\def\s{{\rm s}}
\def\erg{{\rm\,erg}}
\def\cm{{\rm cm}}
\def\km{{\rm\,km}}

\def\gm{{\rm\,g}}

\def\yr{{\rm\,yr}}

\def\pomega{\tilde{\omega}}

\begin{document}

\lefthead{Chiang and Goldreich}
\righthead{Apse Alignment}

\title{Apse Alignment of Narrow Eccentric Planetary Rings}

\author{E.~I.~Chiang and P.~Goldreich}

\affil{California Institute of Technology\\
Pasadena, CA~91125, USA}

\authoremail{echiang@tapir.caltech.edu and pmg@nicholas.gps.caltech.edu}

\begin{abstract}
The boundaries of the Uranian $\epsilon$, $\alpha$, and
$\beta$ rings can be fitted by Keplerian ellipses.
The pair of ellipses that outline a given ring
share a common line of apsides. Apse alignment
is surprising because the quadrupole moment of Uranus
induces differential precession. We propose
that rigid precession is maintained by a balance of
forces due to ring self-gravity, planetary
oblateness, and interparticle collisions.
Collisional impulses play an especially dramatic
role near ring edges. Pressure-induced accelerations
are maximal near edges because there (1) velocity
dispersions are enhanced by resonant satellite perturbations,
and (2) the surface density declines steeply.
Remarkably, collisional forces felt
by material in the last $\sim$100 m of a $\sim$10 km wide ring
can increase equilibrium masses up to a factor of $\sim$100.
New ring surface densities are derived which accord
with Voyager radio measurements. In contrast to previous models,
collisionally modified self-gravity appears to
allow for both negative and positive eccentricity
gradients; why all narrow planetary rings exhibit positive
eccentricity gradients remains an open question.
\end{abstract}


\section{INTRODUCTION}

	Each narrow eccentric ring surrounding Uranus is composed
of particles moving on nested elliptical orbits.
The outer and inner edges of a given ring define ellipses
having semi-major axes $a \pm \Delta a/2$ and eccentricities
$e \pm \Delta e/2$, where $\Delta a \ll a$, $\Delta e \ll e$,
and $e \ll 1$. Observed values of $a$, $e$, $\Delta a$, and $\Delta e$
for the Uranian $\epsilon$, $\alpha$, and $\beta$ rings are listed
in Table \ref{param}.

\placetable{param}
\begin{deluxetable}{ccccc}
\tablewidth{0pc}
\tablecaption{Parameters of Eccentric Uranian
Rings\tablenotemark{a}\label{param}}
\tablehead{
\colhead{Ring}  &  \colhead{$a$(km)}      & \colhead{$\Delta a$(km)} &
\colhead{$e\, (\times 10^3)$}   &  \colhead{$\Delta e\, (\times 10^3)$}    }

\startdata
$\epsilon$ & 51149 & 58.1 & 7.936 & 0.711 \nl
$\alpha$ & 44718 & 7.15 & 0.761 & 0.076 \nl
$\beta$ & 45661 & 8.15 & 0.442 & 0.066 \nl
\tablenotetext{a}{Values taken from Tables I and VII of
\markcite{fetal91}French et al. (1991).}

\enddata
\end{deluxetable}

	Remarkably, the set of ellipses describing an
individual ring share a common line of apsides. Apse alignment
is surprising because the oblateness of Uranus causes orbits of particles
with different semi-major axes to precess differentially.
Timescales for differential precession in the absence
of other forces are extremely short; in the case
of the Uranian $\epsilon$ ring, the inner edge would precess
a full revolution relative to the outer edge in 175 years. Rigid precession
of an eccentric planetary ring has remained a problem in ring dynamics
for over 20 years.

	\markcite{gt79}Goldreich \& Tremaine (1979, hereafter GT) proposed that apse
alignment is maintained by self-gravity. Their theory predicts
that the eccentricity gradient across the ring,

\begin{equation}
q_e \equiv a \frac{\partial e}{\partial a} \, ,
\end{equation}

\ni must be positive. A positive eccentricity gradient
in an apse-aligned ring implies that the ring is narrowest at periapse and
widest
at apoapse. Gravitational forces between particles are therefore greatest
near periapse. Material in the inner half of the ring pulls
radially inward on the outer half at periapse,
generating a differential precession which exactly cancels
that due to planetary oblateness.

	Though the prediction that $q_e > 0$ accords with
observations of all known narrow eccentric rings, the standard
self-gravity model (hereafter SSG) predicts Uranian ring masses
that are too low compared to those inferred
from Voyager radio occultations. Ring masses based on observations
exceed predictions by factors of at least $\sim$3 ($\epsilon$ ring)
to $\sim$50 ($\alpha$ and $\beta$ rings) (\markcite{tetal86}Tyler et al. 1986;
\markcite{g90}Gresh 1990; see also the reviews
by \markcite{espo91}Esposito et al. 1991 and \markcite{fetal91}French et al.
1991).
Low surface densities are
particularly problematic for the $\alpha$ and $\beta$ rings.
With SSG surface densities, torques exerted by inner shepherd satellites
would be insufficiently
strong to confine the $\alpha$ and $\beta$ rings against drag
from the distended exosphere of Uranus (\markcite{gp87}Goldreich \& Porco 1987,
hereafter GP). In addition, as discussed by \markcite{getal95}Graps et al.
(1995),
shapes of the $\epsilon$ ring surface density profiles as
derived from occultation light curves do not accord
with SSG predictions.

	This paper points the way towards resolving
these problems. In \S \ref{qual}, a theory of collisionally
modified self-gravity (hereafter CMSG) is qualitatively described.
A simple quantitative model is set forth in \S \ref{quant}, in which new
surface
density profiles are derived for the $\epsilon$ and $\alpha$ rings that are in
better agreement with observations. In \S \ref{impl}, implications of
our solutions for torque balance, the role of planetary oblateness,
and the value of $q_e$ are discussed. Directions for future research
are summarized in \S \ref{future}.

\section{QUALITATIVE SOLUTION}
\label{qual}

	For simplicity, consider an apse-aligned eccentric ring having
constant, positive $q_e$ across its width. The ring is filled with
spherical particles of internal mass density $\rho$ and radius $r$,
and the ring surface density is given by $\Sigma$. Let $n$ and $\pomega$
be the mean motion and apsidal angle, respectively, of a ring particle.
Subscripts $i$ and $b$ denote quantities evaluated in the ring interior
and near the ring boundary, respectively.
Variables subscripted with $p$ or $s$ are associated with the central planet
or shepherd satellite, respectively, and take their usual meanings. The
dimensionless
strength of the quadrupole moment of the planet is given by $J_2$.
Numerical estimates in this section are made using parameters
appropriate for the $\epsilon$ ring.

	A key ingredient missing in the SSG model is an accounting
for interparticle collisions. Since ring optical depths $\tau$
measured normal to the orbital plane are typically of order unity,
each particle collides, on average, a few times with its neighbors every
orbital period.
Only modest collisional impulses per unit mass and time, of order
$0.1 \, \cm \,\s^{-1}$ per orbit, are required to generate differential
precession rates comparable to those induced by planetary oblateness
(\markcite{gt79}GT).
Velocity dispersions of order $c_i \sim 0.1 \,\cm \,\s^{-1}$ in the ring
interior are
not unreasonable: both the Keplerian shearing velocity across a
particle diameter, $\sim 3 n r$, and the escape velocity
from the particle surface, $\sim r \sqrt{8G\rho}$, are of that
order for the meter-sized bodies that plausibly compose the ring.

	Although a single collision can impart an impulse of dynamically
significant magnitude, multiple collisions experienced by a particle
in the ring interior leave its precession rate largely unaltered.
A particle in the ring interior is struck by its inside neighbors
about as frequently and as forcefully as by its outside ones.
Differential precession across the ring induced by smooth internal
pressure gradients occurs on timescales of order
$2\pi \Sigma n a e / |\nabla P| \sim 2\pi n a e \Delta a / c_i^2 \sim 10^6 \,
(0.1 \,\cm \,\s^{-1} / c_i)^2 \yr$,
much longer than misalignment timescales set by planetary oblateness (cf.
\markcite{gt79}GT).
Here the height-integrated
pressure $P \sim \Sigma c_i^2$ is taken to vary over a lengthscale $\Delta a$.

	Conditions are dramatically different near ring edges.
Pressure-induced accelerations are maximal there because (1) velocity
dispersions are enhanced by resonant satellite perturbations,
and (2) the surface density declines steeply (\markcite{bgt82}Borderies,
Goldreich, \& Tremaine 1982). The velocity dispersion
near the ring boundary could be as high as

\begin{equation}
\label{oom}
c_b \sim \sqrt{\frac{d}{w_r}} \, c_i \sim 3 \; \frac{c_i}{0.1 \,\cm\,\s^{-1}}
\;\cm\,\s^{-1} \, ,
\end{equation}

\ni where $d \sim 10^3 \km$ is the ring-satellite separation, and $w_r$ is the
width
of the annulus perturbed by the satellite. To order-of-magnitude,
the latter is given by $w_r \sim a \, \sqrt{M_s/M_p} \sim 1 \km$,
the distance from the resonant edge at which nested periodic orbits cross.
Equation (\ref{oom}) is derived by equating the rate of energy
dissipation by collisions in the perturbed zone,
$\sim \pi \Sigma a w_r c_b^2 n \tau$, to the rate
of energy deposition by the satellite, $\sim 3 n T d / 2 a \sim 9 \pi \Sigma n
a c_i^2 d / 2$,
where $T$ is the satellite-induced confining torque
whose magnitude equals that of the viscous torque, $\sim 3 \pi \Sigma c_i^2
a^2$,
in steady-state.

	A particle on the ring edge experiences
a radially directed, collisional acceleration

\begin{equation}
C \sim -\frac{\nabla P}{\Sigma} \sim \pm \, \frac{c_b^2}{\lambda} \: \hat{r}
\sim \pm \, c_b \, n \: \hat{r}\;,
\end{equation}

\ni where the upper (lower) sign applies to the outer (inner) ring edge.
Here $P \sim \Sigma \, c_b^2$ is taken to vary
over a radial lengthscale, $\lambda$, of order the local ring thickness,
$c_b / n$. In a $q_e > 0$ ring, collision rates are highest
near periapse. At the periapsis of a ring boundary, the radial
acceleration, $C$, generates a differential precession rate,
$\Delta \langle d\pomega / dt\rangle _C \approx -C/ n a e$,
relative to the precession rate at the ring midline.
This collision-induced rate is greater than the local
differential rate due to planetary oblateness,
${\Delta \langle d\pomega / dt\rangle_O}$, by a substantial factor:

\begin{equation}
\label{bigfactor}
\frac{\Delta \langle d\pomega / dt\rangle_C}{\Delta \langle d\pomega /
dt\rangle_O} \sim \frac{\mp c_b / ae}{\mp (21 \Delta a / 8 a) J_2 n ( R_p/a
)^2} \sim 40 \, \frac{c_b}{1 \,\cm \,\s^{-1}} \, .
\end{equation}

	Self-gravity maintains apse alignment against differential
precession caused by planetary oblateness and interparticle collisions.
For self-gravity to enforce rigid precession near ring edges, surface
densities there must be higher than those predicted by SSG.
At ring boundaries, self-gravitational attraction must balance
the extra repulsive acceleration due to collisions.
To estimate the surface density near the edge, $\Sigma_b$,
equate the collisional
acceleration, $C \sim c_b^2 / \lambda$, to the gravitational acceleration
from a wire of linear mass density $\Sigma _b \lambda$ located a distance
$\lambda$ away:

\begin{equation}
c_b^2 / \lambda \sim 2 G \Sigma _b \; .
\end{equation}

\ni Take $\lambda = c_b / n \sim 50 \,{\rm m}$ to obtain

\begin{equation}
\label{sigedge}
\Sigma _b \sim c_b \,n \, / \, 2 \, G \sim 10^3 \frac{c_b}{1 \,\cm
\,\s^{-1}}\,\gm \,\cm^{-2} \;,
\end{equation}

\ni which is greater than corresponding SSG predictions
by factors $\gtrsim 40$. Equation (\ref{sigedge}) is equivalent to the
condition that
Toomre's Q be of order unity at the edge.

	These endwires of mass $\sim 2\pi a\Sigma_b \lambda = \pi c_b^2 a / G$
constitute new boundary conditions not found in SSG. Gravitational forces
from massive endwires
induce substantial differential precession in the ring interior.
For self-gravity to maintain apse alignment in the interior,
surface densities there must also be greater than those predicted by SSG.

\section{QUANTITATIVE MODEL}
\label{quant}

Divide the region occupied by an apse-aligned, constant $q_e = a\Delta e/
\Delta a$
ring into an even number $N$ of equally spaced intervals. The center of the
$j^{\rm th}$ interval contains an elliptical wire having
mass $m_j$, semi-major axis $a_j = a + [\,j - (N+1) / 2\,]\,\Delta a / N$,
and eccentricity $e_j = e + [\,j - (N+1) / 2\,]\,\Delta e / N$.
Denote by $\Delta_j \langle d\pomega / dt \rangle$
the precession rate of the $j^{\rm th}$ wire relative
to the precession rate of a test particle at the ring midline.
Uniform precession requires

\begin{equation}
\label{nequations}
\Delta_j \langle \frac{d\pomega}{dt} \rangle = \Delta_j (\langle
\frac{d\pomega}{dt} \rangle_O + \langle \frac{d\pomega}{dt} \rangle_G + \langle
\frac{d\pomega}{dt} \rangle_C) = 0 \, .
\end{equation}

\ni Subscripts $O$, $G$, and $C$ denote contributions from
planetary oblateness, self-gravity, and interparticle collisions, respectively.
The first two terms are given by

\begin{equation}
\label{delo}
\Delta_j \langle \frac{d\pomega}{dt} \rangle_O =
-\frac{21}{4}J_2n\left(\frac{R_p}{a}\right)^2\frac{a_j-a}{a}
\end{equation}

\ni and

\begin{equation}
\label{delg}
\Delta_j \langle \frac{d\pomega}{dt} \rangle_G = \frac{q_e H(q_e^2)}{\pi e} n
\frac{a}{M_p} \sum_{k\neq j}\frac{m_k}{a_j - a_k} \, ,
\end{equation}

\ni where

\[
H(q_e^2) \equiv \frac{1-\sqrt{1-q_e^2}}{q_e^2 \sqrt{1-q_e^2}}
\]

\ni (cf. \markcite{gt79}GT).

For $\Delta_j \langle d\pomega / dt \rangle_C$, the
following simplistic prescription is adopted:

\begin{equation}
\label{prescrip}
\Delta_j \langle \frac{d\pomega}{dt} \rangle_C = \left\{ \begin{array}{ll}
	+ [\,q_eH(q_e^2)\:c_b^2/\lambda nae\,]\,(1 - x / \lambda) & \mbox{if $x \equiv
(j-1/2)\,\Delta a / N < \lambda$} \\
	- [\,q_eH(q_e^2)\:c_b^2/\lambda nae\,]\,(1 - y / \lambda) & \mbox{if $y \equiv
(N-j+1/2)\,\Delta a / N < \lambda$} \\
	0 & \mbox{otherwise.}
	\end{array}
\right.
\end{equation}

\ni Thus, $\Delta_j \langle d\pomega / dt \rangle_C$
is non-zero only within intervals $\delta a  = \lambda$
from each edge; there, its magnitude rises
linearly from 0 to $q_e H(q_e^2) c_b^2/\lambda nae$.
As a crude justification for this maximum value,
approximate the collisional acceleration as

\begin{eqnarray}
C & \approx & -\frac{\nabla P}{\Sigma} \nonumber \\
\label{collacc}
  & \approx & \pm \frac{c_b^2}{\lambda (1 - q_e \cos f)} \, \hat{r}
\end{eqnarray}

\ni where the upper (lower) sign applies to the outer (inner) edge.
Here the pressure gradient is taken to vary inversely as the
separation between streamlines.
Insert $C$ into Gauss's perturbation equation for $d\pomega /dt$ and
average over true anomaly:

\begin{eqnarray}
\langle\frac{d\pomega}{dt}\rangle & = & -\frac{1}{\pi n a e}\int_0^{\pi} C \cos
f\,df \nonumber \\
\label{azaverage}
 & = & \mp \, q_e H(q_e^2) \, c_b^2 / \lambda nae \, .
\end{eqnarray}

For the constant $q_e$ ring models presented here,

\begin{equation}
\label{simple}
\lambda = c_b / n \, ,
\end{equation}

\ni so that the only remaining free parameter is $c_b$.
Equation (\ref{simple}) is relaxed for \S \ref{scale}
and \S \ref{magqe}.

Note that this prescription for $\Delta_j \langle d\pomega / dt \rangle_C$
ignores the
decrease in velocity dispersion from $c_b$ at the ring edge to $c_i$ in the
ring interior.
The decline in velocity dispersion occurs over a length scale of order $w_r$.
This length scale is large compared to $c_b/n$ so that the gradient of velocity
dispersion
does not give rise to a significant radial acceleration.

For a given value of $c_b$, equations (\ref{nequations}), (\ref{delo}),
(\ref{delg}),
(\ref{prescrip}), and (\ref{simple}) comprise $N$ equations
in $N$ unknowns $\{m_j\}$.\footnote{Reflection symmetry about the ring midline
reduces the number
of equations necessary to $N/2$. Typically
$N \gtrsim 2000$ wires are needed to converge to within 10\% of the solution
for
$N \rightarrow \infty$.}
Solutions for surface density profiles at quadrature for various values of
$c_b$ are
displayed in Figure 1, for parameters appropriate to the $\epsilon$ and
$\alpha$
rings; models for the $\beta$ ring are nearly identical to those of
the $\alpha$ ring. In CMSG models, higher surface densities near ring
edges are evident, as are higher total ring masses.

\begin{figure}
\plotone{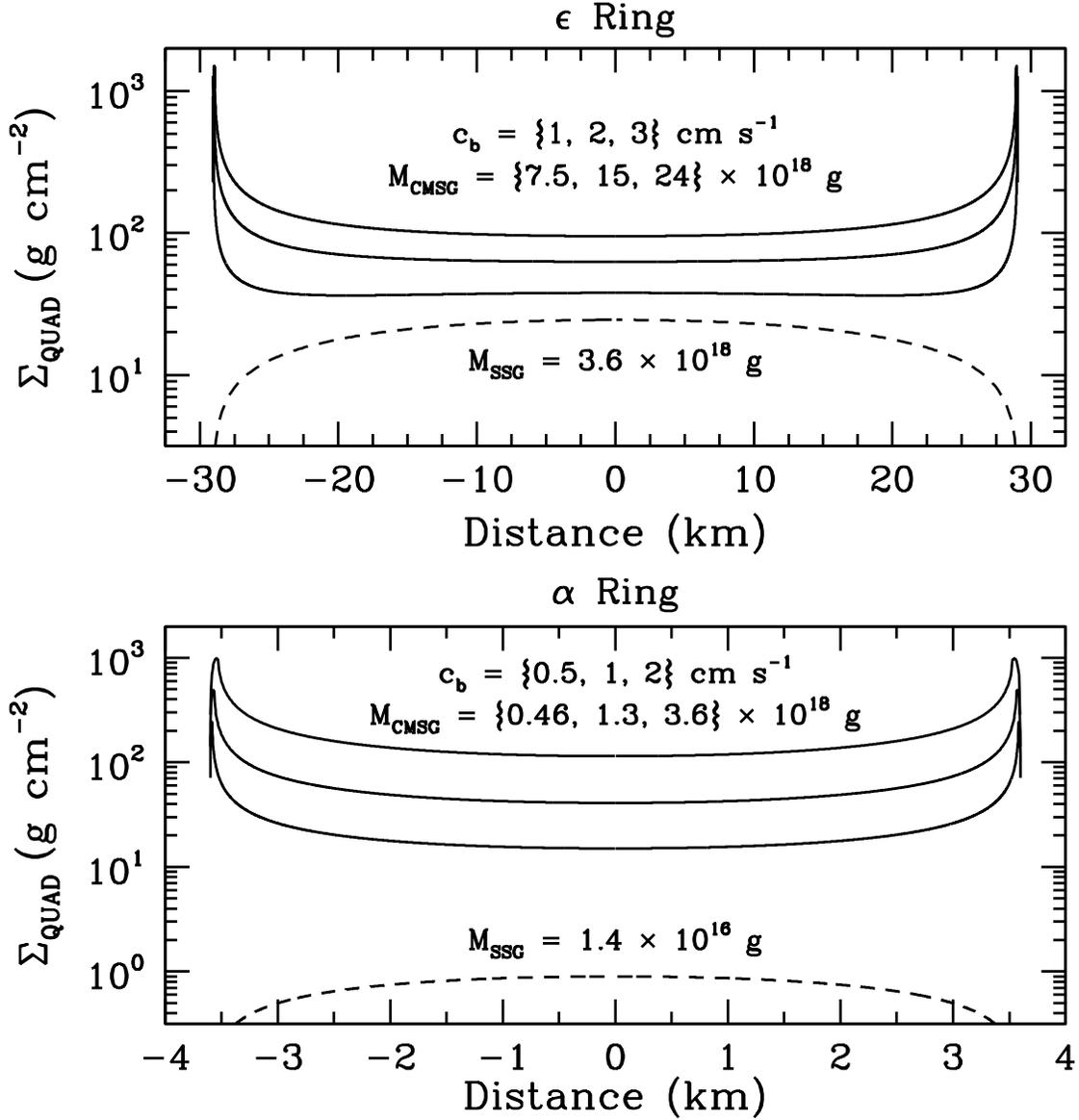}
\caption{Surface density profiles at quadrature for rings $\epsilon$
and $\alpha$ in CMSG ($c_b \neq 0$, solid line) vs.~SSG ($c_b = 0$, dashed
line) models. When the total CMSG ring mass, $M_{\rm{CMSG}}$, greatly
exceeds the total SSG ring mass, $M_{\rm{SSG}}$, it is found empirically that
$M_{\rm{CMSG}} \propto c_b^2 / \lambda^{1/2} \propto c_b^{3/2}$.
\label{surf}}
\end{figure}

\placefigure{surf}

\section{DISCUSSION}
\label{impl}

\subsection{Surface Density Profiles and Torque Balance}

Simple CMSG models, while not fully realistic,
demonstrate the existence of a new class of self-gravity solution,
that obtained by accounting for the modification of ring boundary
conditions by interparticle collisions. Remarkably, forces felt
by material in the last $\sim$100 m of a $\sim$10 km wide ring
can increase equilibrium masses by factors up to 100.

Large S-band opacities measured by Voyager, which are
incompatible with SSG surface densities
(see, e.g., the review by \markcite{espo91}Esposito et al. 1991),
can be reconciled with average CMSG surface densities of $\sim$75--100$\gm
\,\cm^{-2}$
for the $\epsilon$, $\alpha$,
and $\beta$ rings. Moreover, CMSG models predict that surface densities near
ring edges are
higher than those in the interior. This behavior is
reminiscent of the ``double-dip'' structure seen in occultation
light curves for the $\epsilon$ and $\alpha$ rings (see, e.g., the review by
\markcite{fetal91}French et al. 1991).

Greater ring masses as implied by CMSG resolve problems
associated with exospheric drag that were pointed out by \markcite{gp87}GP
for rings $\alpha$ and $\beta$. For the remainder
of this subsection, numerical estimates will be made
for the $\alpha$ ring; similar conclusions hold for the $\beta$ and $\epsilon$
rings. Surface densities are scaled
to a typical CMSG value in the ring interior of $\Sigma = 75 \gm \,\cm^{-2}$.
An inner shepherd satellite
exerts a repulsive, non-linear torque at
first-order Lindblad resonances of magnitude

\begin{equation}
T^L_{\rm nl} \approx \frac{10 \rho_s R_s^3 \Sigma^2 n^2 a^7}{M_p^2 d} \approx 6
\times 10^{17} \left( \frac{\Sigma}{75 \gm \,\cm^{-2}} \right)^2
\left( \frac{R_s}{10 \km} \right)^3 \left( \frac{\rho_s}{1.5 \gm \,\cm^{-3}}
\right)
\left( \frac{500 \km}{d} \right) \erg \;,
\end{equation}

\ni where the satellite radius, $R_s$, is scaled to the Voyager upper limit of
10 km (\markcite{setal86}Smith et al. 1986).
The shepherding torque exceeds the magnitude of the
drag torque exerted by the Uranian exosphere,

\begin{equation}
\label{atm}
T_{d} \approx -4\pi m_H n_H v_T n a^3 \Delta a \approx -4 \times 10^{16} \left(
\frac{n_H}{10^3 \,\cm^{-3}} \right) \erg \, .
\end{equation}

\ni Here $n_H = 7 \times 10^{-6} \, e^{32.4 \, R_p / a} \,\cm^{-3}$ is
the number density of hydrogen atoms of mass $m_H$ in the exosphere, and $v_T
\approx 1 \km \,\s^{-1}$ is
their thermal speed normal to the orbital plane (\markcite{betal86}Broadfoot
et al. 1986). That $T^L_{\rm{nl}} > |T_d|$ ensures that the inner shepherd
prevents ring particles from spiraling in towards Uranus.

Estimates of viscous torques $T_v$ also require revision.
For a ring undergoing Keplerian shear, with minimum kinematic
viscosity $n (\Sigma/\rho)^2$, the viscous torque is given by

\begin{equation}
T_v \approx \frac{3\pi n^2 \Sigma^3 a^2}{\rho^2} \approx 2.5 \times 10^{18}
\left(  \frac{\Sigma}{75 \gm \,\cm^{-2}} \right)^3 \left( \frac{1.5 \gm
\,\cm^{-3}}{\rho} \right)^2 \erg
\end{equation}

\ni (\markcite{gp87}GP). That $T_v \gg |T_d|$ ensures that ring
particles on the outer edge press against the inner Lindblad
resonance established by the outer shepherd.

Conclusions drawn from comparisons between $T^L_{\rm{nl}}$ and $T_v$
are on less sure footing. For the choice of scaling parameters, the latter
exceeds the former, contrary to the requirement of
the standard theory of shepherding that the torques be equal.
This might be construed as evidence that the angular momentum
luminosity in the ring interior is reduced below $T_v$ by the
non-Keplerian shear associated with a non-zero $q_e$
(\markcite{bgt82}Borderies,
Goldreich, \& Tremaine 1982; \markcite{gp87}GP).
However, the numerical estimates for the two torques
differ only by a factor of a few. The shepherding torque
should be evaluated using surface densities near the edge,
which CMSG predicts are higher than those in the interior;
this would increase the estimate of $T^L_{\rm{nl}}$.
Uncertainties in the choice of parameters preclude drawing
any conclusion other than that these torques are of the
same order of magnitude.

\subsection{Relative Importance of Planetary Oblateness}

\subsubsection{$J_2 = 0$ vs. $J_2 \neq 0$}
What does CMSG predict if $J_2 = 0$? Figure 2a displays the answer
for the $\epsilon$ ring, for $c_b = 2$ and 3 $\cm \,\s^{-1}$. In contrast to
SSG,
a non-vanishing equilibrium surface density does not require a finite
planetary oblateness; self-gravity can be balanced entirely by
collisional pressure gradients.
For the $\alpha$ and $\beta$ rings,
solutions with and without $J_2$ are practically indistinguishable
for $c_b \geq 0.5 \,\cm \,\s^{-1}$. The influence of $J_2$ on the
equilibrium solution diminishes as $\Delta a$
decreases or as $c_b$ increases.

\begin{figure}
\plotone{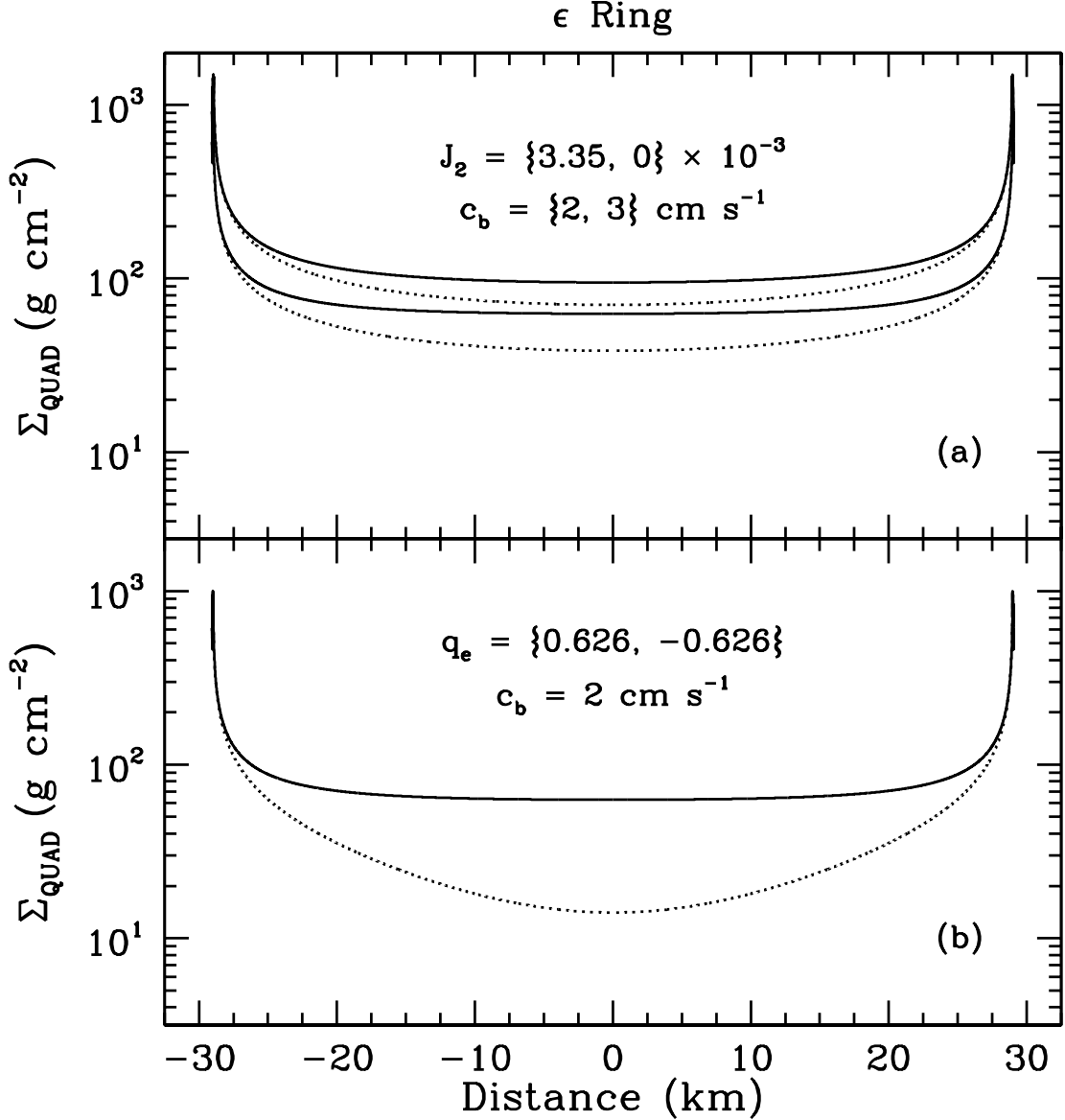}
\caption{(a) CMSG $\epsilon$ ring models for which
$J_2$ is reduced from its nominal value of $3.35 \times 10^{-3}$
{\protect \markcite{en84}}(Elliot \& Nicholson 1984)
(solid line) to 0 (dotted line). As $c_b$ is increased from $2 \, \cm
\,\s^{-1}$ (lower
two curves) to $3 \, \cm \,\s^{-1}$ (upper two curves), the influence of $J_2$
diminishes.
For the $\alpha$ and $\beta$ rings, CMSG models with and without $J_2$
are practically indistinguishable for $c_b \geq 0.5 \,\cm \,\s^{-1}$ (data not
shown).
(b) CMSG $\epsilon$ ring models for which $q_e = \pm \, 0.626$. Contrary to
SSG models, a positive $q_e$ is not required to obtain an equilibrium
solution.}
\label{j2}
\end{figure}

\placefigure{j2}

\subsubsection{Empirical Scaling Relations for $J_2 = 0$}
\label{scale}

For $J_2 = 0$ and fixed ring geometry, the surface density at quadrature
near a given edge scales as

\begin{equation}
\label{scalesurfb}
\Sigma_b \,(0 \leq |x| \lesssim \lambda) = \frac{c_b^2}{G\lambda}
\;f(|x|/\lambda) \, ,
\end{equation}

\ni where $|x|$ measures distance from the edge, $c_b$ and $\lambda$ are the
same
free parameters as in Equation (\ref{prescrip}), and $f$ is a dimensionless
function
of the similarity variable $|x| / \lambda$. Well away from ring edges,
the surface density at quadrature scales as

\begin{equation}
\label{scalesurf}
\Sigma_i \,(|x| \gg \lambda) = \frac{c_b^2}{G\sqrt{\lambda \Delta a}} \;
g(|x|/\Delta a) \, ,
\end{equation}

\ni where $g$ is another dimensionless function. The total ring mass scales as

\begin{equation}
\label{scalemass}
M \sim \frac{c_b^2 a}{G} \sqrt{\Delta a / \lambda} \, .
\end{equation}

\subsection{Value of $q_e$}

\subsubsection{Sign of $q_e$}
Figure 2b displays a CMSG model for the $\epsilon$ ring
obtained by reversing the sign of $q_e$. In contrast to SSG,
a positive eccentricity gradient is not necessary in CMSG
to obtain an equilibrium solution. This resurrects the problem of why all
known eccentric planetary rings, including the Titan and Huygens ringlets
around Saturn, are narrowest at periapse and widest at apoapse.

It is possible that equilibria obtained using $q_e < 0$ are unstable.
To address this issue, a preliminary investigation of ring stability
for an $N=4$ ringlet model has been undertaken. Forces due to pressure
gradients are
included only for the first and fourth ringlets. Collisional accelerations
are treated as if they arise from anti-self-gravity forces (self-gravity
with the sign of the acceleration reversed); i.e., collisional
shear stresses are ignored.
In this crude approximation, equilibria are found to be stable regardless
of the sign of $q_e$; small deviations from equilibrium masses result
in apsidal librations (\markcite{bgt83}Borderies, Goldreich \& Tremaine 1983).
It remains to be seen whether collisional shear stresses alter stability
properties.

Another possibility is that initial conditions set the sign of $q_e$.
If the ring were initially uniform in width
as a function of azimuth, then planetary oblateness
would determine the initial sense of differential precession within the ring.
The resultant narrowing of the ring width near a true anomaly
of $f = -\pi/2$ would cause a positive eccentricity gradient
to grow by self-gravity. Under this hypothesis,
an $N=2$ ringlet model incorporating forces from self-gravity
and planetary oblateness yields the following time evolution for
the apse and eccentricity differences between outer and inner ringlets:

\begin{eqnarray}
\delta \pomega = - A \sin \Omega_{\rm{lib}} t \\
\delta e = A e ( 1 - \cos \Omega_{\rm{lib}} t )
\end{eqnarray}

\ni where $A > 0$ and $\Omega_{\rm{lib}}$ are the amplitude and frequency,
respectively, of libration (cf. \markcite{bgt83}Borderies, Goldreich, \&
Tremaine 1983).
Note that the time-average of $\delta e$ is positive. Inelastic collisions
would damp
librations and the ring would eventually settle into an equilibrium for which
$q_e > 0$.

\subsubsection{Magnitude of $q_e$ Near Ring Boundaries}
\label{magqe}

It has been assumed that the eccentricity gradient, $q_e$, is finite
out to the last $\lambda = c_b/n \sim 50$ meters
of ring material. A finite $q_e$ is necessary to generate
a non-zero azimuthal average of the collisional acceleration
[see equation (\ref{azaverage})]. The simple
quantitative model of \S \ref{quant} employed the
observed value of $q_e$ averaged over the entire
ring width. The true value over the last few hundred
meters of ring material is unknown.

In the case of the best-studied $\epsilon$ ring, \markcite{getal95}Graps et al.
(1995)
combined Voyager photopolarimeter and radio occultation
measurements to infer the eccentricity gradient as a function
of semi-major axis. They found that $q_e$ decreases
over the last $\sim$5 km
from its nearly constant value of $\sim$0.65 in the
interior to $\sim$0.35 near the edge.
The radial resolution of their study was between 1 and 2 km.

A decrease in $q_e$ towards ring boundaries
is theoretically plausible. Distortions
in a circular ring can be described by
the change in separation, $\delta r$, between neighboring
streamlines of the form

\begin{equation}
\delta r \propto \cos m(\phi - \Omega_{\rm pat} t) \, ,
\end{equation}

\ni where $\Omega_{\rm pat}$ is the pattern speed of the distortion
and $m$ is an integer. A constant $q_e$ ring that precesses
rigidly in the quadrupole field of the central planet is
equivalent to a distorted circular ring for which $m = 1$ and $\Omega_{\rm pat}
=
\langle d\pomega/dt \rangle_Q$. Resonant satellite perturbations, which enhance
velocity
dispersions within a distance $w_r \sim 1 \km$ of ring edges, are characterized
by much higher values of $m = 2a/3d \gg 1$ and $\Omega_{\rm pat} = \Omega_s$.
Satellite-induced disturbances might therefore reduce the local value of $q_e$.
A decrease
in $q_e$ over a distance $w_r$ near ring boundaries is roughly equivalent to
setting $\lambda = w_r$
in equation (\ref{prescrip}). By the scaling relations (\ref{scalesurf}) and
(\ref{scalemass}),
this would reduce surface densities and total ring masses shown in Figure
\ref{surf}
by a factor of $\sqrt{w_rn/c_b} \sim 4$.

\section{DIRECTIONS FOR FUTURE RESEARCH}
\label{future}
This work is primarily a demonstration that interparticle
collisions near ring boundaries play a crucial role
in determining ring masses under the self-gravity hypothesis.
The nature of ring boundary conditions has not been calculated
in rigorous detail; instead a prescription motivated
by order-of-magnitude arguments is provided
for collision-induced precession rates.
Numerical simulations incorporating shepherd satellites
will help to determine the actual 3-dimensional collisional stress
tensor and eccentricity gradient everywhere within the ring.

Why all narrow eccentric rings surrounding Uranus and Saturn
are narrowest at periapse and widest at apoapse remains
to be understood. Stability analyses incorporating
collisional shear stresses may reveal that rings having $q_e < 0$ are unstable.
Alternatively, the sign of $q_e$ may be set by initial conditions.
Scenarios for ring formation---e.g., the catastrophic
disruption of a small moon---require further elucidation.

Viscous damping gives rise to small differences
between apsidal angles of neighboring streamlines
(\markcite{bgt83}Borderies, Goldreich, \& Tremaine 1983).
For a given apsidal shift of $\delta \pomega \ll 1$,
the difference between the azimuth of maximum
streamline separation and the azimuth of apoapse is given
by the ``pinch angle'', $\delta \phi =
\arctan (e\delta \pomega / \delta e) \gg \delta \pomega$.
The pinch angles calculated by \markcite{bgt83}Borderies et al. (1983)
for their $N=2$ streamline models of the Uranian and Saturnian ringlets
are suspect, however, because they neglect the boundary effects
highlighted in the present work. A careful calculation
of $\delta \phi (a)$ that incorporates viscous drag and the
global effects of resonant forcing by shepherd satellites
has yet to be performed. Upcoming observations of narrow
Saturnian ringlets by the Cassini Orbiter might test
the predictions of such a calculation, thereby
furnishing a powerful diagnostic of stresses within ringlets.

\acknowledgments
Financial support for this research was provided by NSF grant
94-14232 and by a Caltech Kingsley Foundation Fellowship
held by E.C.


\begin{references}
\reference{bgt82} Borderies, N., Goldreich, P., \& Tremaine, S. 1982, \nat,
299, 209
\reference{bgt83} Borderies, N., Goldreich, P., \& Tremaine, S. 1983, \aj, 88,
1560
\reference{betal86} Broadfoot, A.L. et al. 1986, Science, 233, 74
\reference{en84} Elliot, J.L., \& Nicholson, P.D. 1984, in Planetary Rings, ed.
R. Greenberg \& A. Brahic (University of Arizona Press), 25
\reference{espo91} Esposito, L.W., Brahic, A., Burns, J.A., \& Marouf, E.A.
1991, in Uranus, ed. J.T. Bergstrahl, E.D. Miner, \& M.S. Matthews (University
of Arizona Press), 410
\reference{fetal91} French, R.G., Nicholson, P.D., Porco, C.C., \& Marouf, E.A.
1991, in Uranus, ed. J.T. Bergstrahl, E.D. Miner, \& M.S. Matthews (University
of Arizona Press), 327
\reference{gp87} Goldreich, P., \& Porco, C.C. 1987, \aj, 93, 730 (GP)
\reference{gt79} Goldreich, P., \& Tremaine, S. 1979, \aj, 84, 1638 (GT)
\reference{getal95} Graps, A.L., Showalter, M.R., Lissauer, J.J., \& Kary, D.M.
1995, \aj, 109, 226
\reference{g90} Gresh, D. L. 1990, Voyager Radio Occultation by the Uranian
Rings: Structure,
Dynamics, and Particle Sizes. Ph. D. Thesis, Stanford University
\reference{setal86} Smith, B.A. et al. 1986, Science, 233, 43
\reference{tetal86} Tyler, G.L. et al. 1986, Science, 233, 79
\end{references}
\end{document}